\documentclass[10pt,prc,twocolumn,nofootinbib,noshowkeys,noshowpacs,mgroupedaddress,floatfix]{revtex4-1}

\usepackage{amsmath,amsfonts,amsthm,amssymb}
\usepackage[dvips]{graphics,graphicx}
\usepackage[usenames,dvipsnames]{color}
\definecolor{darkblue}{RGB}{0,0,196}
\definecolor{darkgreen}{RGB}{0,120,0}
\usepackage[colorlinks=true,linktocpage=true,linkcolor=darkblue,citecolor=red,urlcolor=darkblue]{hyperref}
\usepackage{cancel}
\usepackage{bbold}
\usepackage{multirow}
\usepackage{longtable}
\usepackage{color}
\usepackage[normalem]{ulem}
\usepackage{hyperref}
\usepackage{bigints}
\usepackage{xparse}
\usepackage{physics}
\usepackage{verbatim}
\usepackage{minibox}
\usepackage{comment}
\usepackage{appendix}
\usepackage{marginnote}
\usepackage{graphicx}
\usepackage[nice]{nicefrac}
\def\HP{\hphantom{\alpha}} 
\usepackage{amsmath}

\def\be{\begin{equation}}
	\def\ee{\end{equation}}
\newcommand{\bel}[1]{\begin{eqnarray}\label{#1}}
	\newcommand{\eel}{\end{eqnarray}}
\def\barr{\begin{array}}
	\def\earr{\end{array}}
\def\beq{\begin{eqnarray}}
	\def\eeq{\end{eqnarray}}
\def\bfig{\begin{figure}}
	\def\efig{\end{figure}}

\newcommand{\nn}{\nonumber}

\newcommand{\onehalf}{{\nicefrac{1}{2}}}

\newcommand{\p}{\partial}

\newcommand{\rf}[1]{Eq.~(\ref{#1})}
\newcommand{\rfm}[1]{Eqs.~(\ref{#1})}
\newcommand{\rftwo}[2]{Eqs.~(\ref{#1})~and~(\ref{#2})}
\newcommand{\rfmtwo}[2]{Eqs.~(\ref{#1})-(\ref{#2})}
\newcommand{\rfn}[1]{(\ref{#1})}

\def\a{\alpha}
\def\b{\beta}
\def\g{\gamma}
\def\d{\delta}

\def\LR{\left(} 
\def\RR{\right)}

\def\HP{\hphantom{\alpha}} 

\newcommand{\sh}[1]{\sinh#1}
\newcommand{\ch}[1]{\cosh#1}

\def\half{\frac{1}{2}}

\def\GLW{{\rm GLW}}

\newcommand{\lab}[1]{\label{#1}}
\def\nn{\nonumber}
\newcommand{\olra}{\overleftrightarrow}
\newcommand{\refb}[1]{(\ref{#1})}

\def\cA{{\cal A}}
\def\cB{{\cal B}}
\def\cC{{\cal C}}
\def\cD{{\cal D}}
\def\cN{{\cal N}}
\def\cE{{\cal E}}
\def\cP{{\cal P}}
\def\cS{{\cal S}}
\def\cNN{{\cal N}_{(0)}}
\def\cEN{{\cal E}_{(0)}}
\def\cPN{{\cal P}_{(0)}}
\def\cSN{{\cal S}_{(0)}}

\def\be{\begin{equation}}
\def\ee{\end{equation}}
\def\ba{\begin{eqnarray}}
\def\ea{\end{eqnarray}}   

\def\a{\alpha}
\def\b{\beta}
\def\g{\gamma}
\def\d{\delta}

\def\LR{\left(} 
\def\RR{\right)}

\def\half{\frac{1}{2}}

\def\GLW{{\rm GLW}}

\def\half{\frac{1}{2}}
\def\GLW{{\rm GLW}}

\def\n0{n_{(0)}}
\def\e0{\varepsilon_{(0)}}
\def\P0{P_{(0)}}

\begin{document}

\preprint{}

\title{Spin polarization dynamics in the Gubser-expanding background}

\author{Rajeev Singh}
\email{rajeev.singh@ifj.edu.pl}
%
%
\author{Gabriel Sophys}
\email{gabriel.sophys@ifj.edu.pl}
\author{Radoslaw Ryblewski}
\email{radoslaw.ryblewski@ifj.edu.pl}

\affiliation{Institute of Nuclear Physics Polish Academy of Sciences, PL 31-342 Krak\'ow, Poland}
\begin{abstract}
Evolution of spin polarization in the Gubser-expanding conformal perfect-fluid hydrodynamic background is studied. The analysis of the conformal transformation properties of the conservation laws is extended to the case of the angular momentum conservation. The explicit forms of equations of motion for spin components are derived and analysed, and some special solutions are found.
\end{abstract}
\date{\today} 
\maketitle

\section{Introduction}
\label{sec:introduction}
In the last decades relativistic hydrodynamics has proven to be very successful in describing the evolution of the strongly-interacting matter produced in relativistic heavy-ion collisions~\cite{Florkowski:1321594,Gale:2013da,Jeon:2015dfa,Jaiswal:2016hex,Romatschke:2017ejr,Florkowski:2017olj,Alqahtani:2017mhy,Berges:2020fwq}. Recent measurements of spin polarization of $\Lambda$ hyperons  
have shown that the space-time evolution of quantum spin should also be included in this framework in order to properly capture effects seen in the experiment~\cite{Abelev:2007zk,STAR:2017ckg, Adam:2018ivw,Adam:2019srw,Acharya:2019vpe,Kornas:2019,Acharya:2019ryw}.

The first attempt to formulate relativistic hydrodynamics with spin has been done in Ref.~\cite{Florkowski:2017ruc}; see also the follow-up papers~\cite{Florkowski:2017dyn,Florkowski:2018myy,Becattini:2018duy,Florkowski:2018ahw,Florkowski:2019voj,Florkowski:2019qdp,Bhadury:2020puc,Bhadury:2020cop}, as well as reviews \cite{Florkowski:2018fap,Tinti:2020gyh}. In contrast to previous theoretical studies which considered generation of spin polarization of matter at freeze-out due to spin-vorticity coupling~\cite{Becattini:2013fla,Becattini:2013vja,Becattini:2016gvu,Karpenko:2016jyx,Xie:2017upb,Sun:2017xhx,Li:2017slc,Becattini:2017gcx,Wei:2018zfb,Xia:2018tes,Sun:2018bjl,Ivanov:2019wzg,Becattini:2019ntv,Zhang:2019xya,Kapusta:2019ktm,Wang:2020pej,Fukushima:2020ucl,Fu:2020oxj,Gao:2020lxh,Gao:2020pfu,Huang:2020dtn} (see also related studies~\cite{Voloshin:2004ha,Betz:2007kg,Avkhadiev:2017fxj,Baznat:2017jfj,Montenegro:2017rbu,Montenegro:2018bcf,Fukushima:2018grm,Prokhorov:2018bql,McInnes:2018pmk,Huang:2018aly,Yang:2018lew,Xia:2019fjf,Li:2019qkf,Kapusta:2019sad,Liu:2019krs,Ivanov:2019ern,Zhao:2019hta,Ambrus:2019ayb,Sheng:2019kmk,Gao:2019znl,Hattori:2019lfp,Prokhorov:2019yft,Prokhorov:2019hif,Freese:2019bhb,Huang:2020wrr,Guo:2020zpa,Prokhorov:2020okl,Becattini:2020qol,Becattini:2020xbh,Ivanov:2020wak,Deng:2020ygd,Hou:2020mqp,Yang:2020hri,Gallegos:2020otk,Kawaguchi:2020kce,Li:2020dwr,Li:2020eon,Hattori:2020gqh,Shi:2020htn,Garbiso:2020puw,Huang:2020kik,Gallegos:2021bzp}), 
the spin hydrodynamic approach has been formulated entirely based on the conservation laws and assumption of local thermal equilibrium. In this case, enforcement of total angular momentum conservation leads to the need for introducing additional Lagrange multipliers which comprise the so-called spin polarization tensor $\omega^{\mu\nu}$ that in the case of global equilibrium with rigid rotation reduces to the thermal vorticity tensor $\varpi_{\mu \nu} = -\frac{1}{2} (\p_\mu \b_\nu-\p_\nu \beta_\mu)$ (here $\beta_\mu = U_\mu/T$ is the ratio of the fluid flow vector $U_\mu$ and the local temperature $T$)~\cite{Becattini:2007nd,Becattini:2009wh,Becattini:2015nva}. The resulting spin hydrodynamic framework extends the standard one by adding additional dynamic equations which, in general, have to be solved numerically.

In the present paper we use the formalism of relativistic hydrodynamics with spin proposed in Refs.~\cite{Florkowski:2018fap,Florkowski:2019qdp} to study evolution of spin polarization in the Gubser-expanding perfect-fluid hydrodynamic background~\cite{Gubser:2010ze,Gubser:2010ui}. The current work is an extension of the study presented in Ref.~\cite{Florkowski:2019qdp} where the space-time evolution
of the spin polarization was considered in a boost-invariant and transversely homogeneous background, also known as Bjorken flow. Following other works~\cite{Denicol:2014xca,Nopoush:2014qba,Nopoush:2015yga,Martinez:2017ibh,Chattopadhyay:2018apf,Calzetta:2019dfr,Behtash:2019qtk,Dash:2020zqx,Jiang:2020big,Shokri:2020cxa} we first solve the perfect-fluid hydrodynamical equations using the Gubser symmetry arguments in the de Sitter coordinates and obtain analytical solutions for hydrodynamic variables. Subsequently, we extend the analysis of the properties of the conservation laws with respect to the conformal transformation to the case of the angular momentum conservation. We find that the latter is conformaly invariant only if the spin tensor is antisymmetric in all indices and the particles composing the fluid are massless. As the de Groot - van Leeuwen - van Weert (GLW) spin tensor~\cite{DeGroot:1980dk} considered in this work is not, in general, satisfying these requirements, we relax the constraints related to the symmetry with respect to the special conformal transformations exploiting only the cylindrical symmetry and boost-invariance. Using expressions for temperature, chemical potential and velocity obtained for the Gubser-flowing background, we derive the equations of motion for spin polarization components which, as expected,  exhibit non-trivial dynamics in both de Sitter time and angle, as well as a (small) sensitivity to the mass of the constituent particles. As in the formulation used herein we keep the assumption of small amplitude of polarization effects~\cite{Florkowski:2018fap,Florkowski:2019qdp} the background equations expressing the baryon charge and energy-momentum conservation decouple completely from the angular momentum conservation, the background solutions are not spoiled by the breaking of the symmetry at the level of angular momentum conservation.  For the special case of massless particles we find a set of special solutions which, in general, have a power-like dependence on the temperature of the system. Finally, we also present some full numerical solutions with finite masses in Milne space-time. The solutions obtained in this work may be used for describing the collisions of systems with initial spin polarization as well as for testing the full 3+1D geometry codes.

\smallskip

%
%
\smallskip

The structure of the paper is as follows: We start by briefly reviewing the framework of relativistic hydrodynamics with spin  for polarized systems of spin -$\onehalf$ particles in Sec.~\ref{sec:perfect}. In Sec.~\ref{sec:implement} we introduce the boost-invariance and cylindrical symmetry, which is followed in Sec. \ref{sec:gubser} by the details about Gubser symmetry, Weyl rescaling of various thermodynamical and hydrodynamical quantities, and the information about the de Sitter basis vectors. Conformal transformation of the conservation equations is studied in Sec. \ref{sec:invariance}. In Sec. \ref{sec:dynamics} we derive the Gubser-symmetric evolution equations for the background which we then use to finding the evolution equations of spin polarization components in the de Sitter spacetime. The resulting equations of motion are studied both analytically and numerically and transformed back to Milne spacetime. We conclude and summarize in Sec. \ref{sec:summary}. Some additional information and details are given in the appendices.
\section{Conventions}
\label{sec:Conv}
We use the following shorthand notation for the scalar product between two four-vectors  $a_\mu b^\mu \equiv a \cdot b$. For the Levi-Civita tensor $\epsilon^{\a\b\g\d}$, the sign convention used throughout the paper is $\epsilon^{0123}=-\epsilon_{0123}=1$. Throughout the text we use natural units with $c = \hbar = k_B=1$. The anti-symmetrization of arbitrary rank two tensor $A$ is denoted as $A_{[\mu \nu]} =   \frac{1}{2!}\left(A_{\mu\nu} - A_{\nu\mu} \right)$. Also, we are using ``mostly plus'' metric signature.
%
\section{Perfect fluid hydrodynamics for systems of particles with spin 1/2}
\label{sec:perfect}
%
For reader's convenience in this section we briefly review the recently developed formalism of relativistic perfect fluid hydrodynamics for polarized systems of particles with spin $\onehalf$ \cite{Florkowski:2019qdp}. We use the approximation of small polarization which, in the leading order, leads to decoupling of the background hydrodynamic equations for velocity, temperature and chemical potential (given by the standard net baryon number and energy-momentum conservation laws) from the dynamics of spin degrees of freedom (resulting from the conservation of angular momentum). 
%
\subsection{Conservation of net baryon number}
\label{sec:barcons}
%
The conservation law of net baryon number can be expressed in the following way
\bel{eq:Ncon}
d_\a N^\a(x)  = 0\, ,
\eel
where, in the case of no dissipative effects, the net baryon current $N^\a$ reads
\bel{eq:N}
N^\a = \cN U^\a\, ,
\eel
with the net baryon density being given by
\bel{eq:calN}
\cN=4  \sh\alpha \,\, \cNN\, .
\eel
The fluid four-velocity $U^\a$ in \rf{eq:N} is normalized as $U \cdot U \!=\!-1$ and $d$ denotes the covariant derivative (see App.~\ref{app:covderiv} for more information on the definitions of covariant derivative).
The quantity $\alpha$ is the ratio of baryon chemical potential $\mu$ and temperature $T$, $\alpha\equiv\mu/T$. For an ideal relativistic gas of classical massive particles (and antiparticles) the auxiliary number density $\cNN$ is given by the well known expression \cite{Florkowski:1321594}
\bel{eq:calNN}
\cNN &=&    k \, T^{3} z^2 K_{2}\left( z \right)\, , 
\eeq
where $z\!\equiv\!m/T$ with $m$ being the mass of the particles (and antiparticles), $k\!\equiv\!1/(2\pi^2)$ and $K_{n}$ denotes the $n$-th modified Bessel function of the second kind.
%
\subsection{Conservation of energy and linear momentum}
\label{sec:enecons}
%
The conservation of energy and linear momentum is given by the expression
\bel{eq:Tcon}
d_\a T^{\a\b}(x) = 0\, ,
\eel
where for perfect fluid the energy-momentum tensor $T^{\a\b}$ takes the following standard form
\bel{eq:T}
T^{\a\b} &=& \cE \, U^\a U^\b + \cP \Delta^{\a\b}\, ,
\eel
with $\Delta^{\a\b} \equiv g^{\a\b} + U^\a U^\b$ being the spatial projection operator orthogonal to the fluid flow vector and the energy density and pressure being defined as 
\beq
\cE&=&4 \ch\alpha \,\, \cEN\, ,
\label{eq:calE} \\
\cP&=&4 \ch \alpha \,\, \cPN\, ,
\label{eq:calP}
\eeq
respectively. In the case of ideal relativistic gas of spinless and neutral massive Boltzmann particles the auxiliary energy density and pressure is \cite{Florkowski:1321594}
\beq 
\cEN &=& k  \, T^4 \, z^2
 \Big[z  K_{1} \left( z \right) + 3 K_{2}\left( z \right) \Big]\,,  \label{eq:E0}\\
\cPN &=&   T  \cNN  \, ,\label{eq:P0} 
\eeq
respectively. The thermodynamic quantities defined above satisfy the standard thermodynamic relations, namely 
\beq 
\cE + \cP = T \cS + \mu \cN\, , \label{eq:therm}\\
\cEN + \cPN = T \cSN\, ,
\label{eq:therm0}
\eeq
with $\cS = \LR\frac{\p \cP}{\p T}\RR_{\mu}$, $\cN = \LR\frac{\p \cP}{\p \mu}\RR_{T}$, and $\cSN =\frac{\p \cPN}{\p T}$.
%
\subsection{Conservation of angular momentum}
\label{sec:angcons}
Total angular momentum may be expressed as a sum of orbital angular momentum part $L^{\a,\b\g}$ and spin angular momentum part $S^{\a,\b\g}$ as follows
\beq
J^{\a,\b\g}&=&L^{\a,\b\g} +S^{\a,\b\g}=x^{\b} T^{\a\g} - x^{\g} T^{\a\b} + S^{\a,\b\g}\, ,\nn\\ 
\label{eq:L}
\eeq
where in the second equality we expressed the orbital part in terms of the energy-momentum tensor.
The conservation of total angular momentum is given by the expression
\beq
d_\a J^{\a,\b\g}=d_\a S^{\a,\b\g}+ 2 T^{[\b\g]}= 0\, .
\label{eq:Scon}
\eeq
%
Since the energy-momentum tensor used herein is symmetric $T^{[\b\g]}= 0$, see Eq.~(\ref{eq:T}), the conservation of the total angular momentum implies separate conservation of its spin part.

The spin tensor in Eq.~(\ref{eq:Scon}) is given by the following expression \cite{Florkowski:2019qdp} 
\beq
S^{\a,\b\g}
&=&    \cC \,U^\a \omega^{\b\g}  +  S^{\a, \b\g}_{\Delta}  ,
\label{eq:S}
\eeq
where the auxiliary spin tensor is defined as
\beq
S^{\a, \b\g}_{\Delta} 
&=&  \cA \, U^\a U^\d U^{[\b} \omega^{\g]}_{\HP\d} \label{eq:SDelta} \\
&& \hspace{-0.5cm} + \, \cB  \, \Big( 
U^{[\b} \Delta^{\a\d} \omega^{\g]}_{\HP\d}
+ U^\a \Delta^{\d[\b} \omega^{\g]}_{\HP\d}
+ U^\d \Delta^{\a[\b} \omega^{\g]}_{\HP\d}\Big)\, ,
\nn
\eeq
with the quantity $\omega^{\a\b}(x)$ being the spin polarization tensor (to be discussed in the next section). The thermodynamic quantities $ \cA$, $\cB$ and $\cC$ in \rftwo{eq:S}{eq:SDelta} are 
\ba
\cA &=& 2 {\cal C} -3 {\cal B}\, ,
\label{eq:A} \\
\cB &=&-\frac{\cE+\cP}{2 T z^2 }\, , \label{eq:B}\\
\cC&=& \frac{\cP}{4T}\, .
\label{eq:C}
\ea 
In what follows here on-wards we will refer
\rftwo{eq:S}{eq:SDelta} as the de Groot - van Leeuwen - van Weert (GLW) spin tensor~\cite{DeGroot:1980dk}.
\subsection{Spin polarization tensor}
%
The spin polarization tensor $\omega^{\a\b}$ is an asymmetric rank-two tensor which can be decomposed with respect to the fluid four-velocity in the following way
\beq
\omega^{\a\b} &=& \kappa^\a U^\b - \kappa^\b U^\a + \epsilon^{\a\b\g\d} U_\g \omega_{\d}\, .
\label{spinpol}
\eeq
Any part of the auxiliary four-vectors $\kappa^{\a}$ and $\omega^{\a}$ parallel to  $U^{\a}$ does not contribute to the right-hand side of~Eq.~(\ref{spinpol}) hence without the loss of generality we can assume that $\kappa^{\a}$ and $\omega^{\a}$  satisfy the following orthogonality conditions
\beq
\kappa\cdot U = 0\, , \quad \omega \cdot U = 0  \label{ko_ortho}\, .
\eeq
Using above conditions, $\kappa_\a$ and $\omega_\a$ can be expressed in terms of spin polarization tensor $\omega_{\a\b}$ as follows
\beq
\kappa_\a= \omega_{\a\b} U^\b\, , \quad \omega_\a = \half \epsilon_{\a\b\g\d} \omega^{\b\g} U^\d\, . \lab{eq:kappaomega}
\eeq
%
%
\section{Implementation of boost invariance and cylindrical symmetry}
\label{sec:implement}
%
Let us consider central high-energy heavy-ion collisions which are boost-invariant and cylindrically-symmetric with respect to the beam ($z$) axis.  Their dynamics is most conveniently described in the polar hyperbolic coordinates $x^\mu = (\tau,r,\phi,\eta)$ where the line element reads
\beq
ds^2=-d\tau^2+dr^2 + r^2d\phi^2 + \tau^2d\eta^2,
\eeq
with $\tau =\sqrt{t^2 - z^2}$ being the longitudinal proper time, $\eta ={\rm tanh}^{-1}(z/t)$ being the longitudinal spacetime rapidity,
and $r=\sqrt{x^{2}{+}y^{2}}$ 
and $\phi ={\rm tan}^{-1}(y/x)$ denoting the radial distance and the azimuthal angle,  respectively, parameterizing the
transverse plane perpendicular to the beam direction.

For convenience we introduce the following orthogonal four-vector basis in the laboratory frame 
\ba 
U^\mu &=& (\ch \vartheta,\sinh \vartheta,0,0)\, , \nonumber \\
R^\mu &=& (\sh \vartheta,\ch \vartheta ,0,0)\, , \nonumber \\
\Phi^\mu &=& (0,0,1/r,0)\, ,\nonumber \\
Z^\mu &=& (0,0,0,1/\tau)\, , 
\label{eq:pmbasis} 
\ea
which allows us to express the polar-hyperbolic metric tensor, \mbox{$g^{\mu\nu}={\rm diag}(-1,1,1/ r^2,1/\tau^2)$}, in the following form
\ba
g^{\mu\nu}=-U^\mu U^\nu+R^\mu R^\nu+\Phi^\mu \Phi^\nu+Z^\mu Z^\nu \, .
\label{eq:pmmetric}
\ea
It is straightforward to check that the basis vectors \refb{eq:pmbasis} satisfy the following normalization conditions
\ba
&&U\cdot U  =-1\,, \quad
R \cdot R =1\, ,\quad\\\,\,
&&\Phi\cdot \Phi =1\, ,\,\,\quad \quad
Z\cdot Z =1\, ,
\label{eq:relations}
\ea 
while all mixed scalar products vanish. One can convince himself that in the local rest frame $U$, $R$, $\Phi$ and $Z$ are the unit vectors pointing in the $\tau$, $r$, $\phi$ and $\eta$ directions, respectively.

Using the basis vectors \refb{eq:pmbasis} and orthogonality conditions \refb{ko_ortho} four-vectors  $\kappa^\a$ and $\omega^\a$ may be decomposed as follows
\beq
\kappa^\a &=&  a_R R^\a + a_\Phi \Phi^\a + a_Z Z^\a, \lab{eq:k_decom}\\
\omega^\a &=&  b_R R^\a + b_\Phi \Phi^\a + b_Z Z^\a, \lab{eq:o_decom}
\eeq
where $a_i(\tau,r)$ and $b_i(\tau,r)$ are scalar coefficients characterizing the components of spin polarization tensor along the basis vectors.
%
\section{Gubser symmetry and conformal mapping to de Sitter space}
\label{sec:gubser}
%
For boost-invariant and cylindrically symmetric  systems one can construct a nontrivial four-velocity profile in Minkowski space $R^3{\otimes}{R}$ which is invariant with respect to the  $SO(3)_q{\otimes}SO(1,1){\otimes}Z_2$ conformal symmetry group, also known as ``Gubser's symmetry'' \cite{Gubser:2010ze,Gubser:2010ui}. The latter is composed of rotations in the $r-\phi$ plane coupled with two special conformal
transformations parametrized by an arbitrary inverse length scale $q$ ($SO(3)_q$), boosts in the $\eta$ direction ($SO(1,1)$) and reflections with respect to the $r-\phi$ plane ($Z_2$). The procedure of finding the flow pattern invariant with respect to this symmetry, while being  intricate in Minkowski space, is quite transparent when considered in the curved spacetime formed by the product of the three-dimensional de Sitter space and a line, $dS_3 {\otimes} {R}$ -- hereafter we will shortly refer to it as ``de Sitter space''. The conformal map used to transform from one to the other constitutes of the Weyl rescaling of the line element 
\beq
ds^2 \to \frac{ds^2}{\Omega^2}=\frac{-d\tau^2+dr^2 + r^2d\phi^2}{\tau^2} + d\eta^2\, ,
\label{eq:deSitter-WR}
\eeq
where the conformal factor is $\Omega=\tau$,
combined with the change from polar Milne coordinates $x^\mu=(\tau,\,r,\,\phi,\,\eta)$ to de Sitter  coordinates $\hat{x}^\mu=(\rho,\,\theta,\,\phi,\,\eta)$ using
\ba 
\sinh{\rho(\tau,r)} &=&  - \frac{1-(q \tau)^2+(q r)^2}{2~q{\tau}}\, ,
\label{eq:desitter1} \\
\tan{\theta(\tau,r)} &=& \frac{2~qr}{1+(q \tau)^2-(q r)^2}\, .
\label{eq:desitter2}
\ea
The resulting rescaled line element of the de Sitter spacetime reads
\beq
d\hat{s}^2 = -d\rho^2 + \cosh^2\!\rho (d\theta^2 + \sin^2\!\theta ~d\phi^2) + d\eta^2\,,
\label{eq:deSitter-ds}
\eeq
with the metric \mbox{$
\hat{g}_{\mu\nu} = {\rm diag}(-1,\, \cosh^2\!\rho,\, \cosh^2\!\rho\,  \sin^2\!\theta ,\, 1)
$} (hereafter all quantities defined in the de Sitter space will be denoted with a hat). 
Clearly, based on \rf{eq:deSitter-ds} we observe that the Weyl rescaling (\ref{eq:deSitter-WR}) combined with passing to the  standard global coordinates on $dS_3$ (\ref{eq:desitter1})-(\ref{eq:desitter2}) promotes the ($SO(3)_q$) conformal isometry to a manifest isometry ($SO(3)$) in $(\theta,\phi)$.

In general, for a system to respect conformal symmetry, its dynamics should be invariant under Weyl rescaling \cite{Baier:2007ix,Bhattacharyya:2007vs,Loganayagam:2008is,Gubser:2010ze,Gubser:2010ui}. It implies that the $(m,n)$-type tensors (including scalars with $(m,n)=(0,0)$) transform homogeneously, namely
\ba
 A^{\mu_1 ...\mu_m}_{\nu_1 ...\nu_n}(x)\,\rightarrow\,\Omega^{\Delta_A}A^{\mu_1 ...\mu_m}_{\nu_1 ...\nu_n}(x)\, ,
\label{eq:weyl-rescaling}
\ea
where $\Omega \equiv e^{-\varphi (x)}$ with $\varphi(x)$ being function of space-time coordinates and $\Delta_A = [A]+m-n$ is the conformal weight of the quantity $A$,
where $[A]$ is its mass dimension, and $m$ and $n$ being the number of contravariant and covariant indices, respectively.
For instance, the metric tensor $g_{\mu\nu}$ is a rank-two $(0,2)$ dimensionless tensor, $[g_{\mu\nu}]=0$, for which one finds $\Delta_{g_{\mu\nu}}\!=\!-2$. Hence, $g_{\mu\nu}$  transforms under Weyl rescaling as follows \cite{Baier:2007ix,Gubser:2010ui}
\ba 
g_{\mu\nu} \rightarrow \Omega^{-2}\,g_{\mu\nu}\, .
\label{eq:g-weyl}
\ea
Using rules for general coordinate transformations of tensors and Eq.~(\ref{eq:g-weyl}) the relation between $R^3{\otimes}{R}$  metric and $dS_3{\otimes}{R}$  metric is
\ba
\hat{g}_{\mu\nu}=\frac{1}{\tau^2}\frac{\partial x^\alpha}{\partial \hat{x}^\mu}\frac{\partial x^\beta}{\partial\hat{x}^\nu}g_{\alpha\beta}\, .
\ea
Using the information that $\Delta_{g_{\mu\nu}}\!=\!-2$ and  the unit norm constraint $g_{\mu\nu} U^\mu U^\nu=-1$, one obtains $\Delta_{U^\mu}=1$ which results in the transformation rule
\beq
\hat{U}_{\nu}=\frac{1}{\tau} \frac{\partial x^{\mu}}{\partial \hat{x}^{\nu}} U_{\mu}\,.
\label{eq:Urel}
\eeq
Using Eq.~(\ref{eq:Urel}) one can show that the four-velocity profile from Eqs.~(\ref{eq:pmbasis}) in the de Sitter space is static,
\ba 
\hat{U}^\mu &=& (1,\,0,\,0,\,0)\, , \nonumber 
\ea
meaning it is invariant with respect to the Gubser symmetry group, provided the transverse rapidity profile has the form \cite{Gubser:2010ui,Gubser:2010ze} 
\be 
\vartheta(\tau,r)= \tanh^{-1}\left(\frac{2 \,q \tau \,  q r}{1+(q \tau)^2+(q r)^2}\right)\, .
\label{eq:thetaperp}
\ee 
In the similar manner the  remaining basis vectors (\ref{eq:pmbasis}) in the de Sitter space are
\ba 
\hat{R}^\mu &=& (0,\,(\cosh\rho)^{-1},\,0,\,0)\, , \nonumber \\
\hat{\Phi}^\mu &=& (0,\,0,\,(\cosh\rho \sin\theta )^{-1},\,0)\, , \nonumber \\
\hat{Z}^\mu &=& (0,\,0,\,0,\,1)\, .
\label{eq:desitter-4vectors}
\ea
The metric $\hat{g}^{\mu\nu}$ can be expressed as
\ba
\hat{g}^{\mu\nu}=-\hat{U}^\mu \hat{U}^\nu+\hat{R}^\mu \hat{R}^\nu+\hat{\Phi}^\mu \hat{\Phi}^\nu+\hat{Z}^\mu \hat{Z}^\nu \, ,
\ea
while the determinant of $\hat{g}_{\mu\nu}$ is
\ba
\hat{g} \equiv \det(\hat{g}_{\mu\nu}) = -\cosh^4\!\rho \sin^2\!\theta\, .
\label{eq:g}
\ea
Using the fact that the energy density \rf{eq:calE} and pressure \rf{eq:calP} have the mass dimension $[\cE]\equiv[\cP]=4$, their conformal weight is $\Delta_{\cE}=\Delta_{\cP}=4$.
In a similar way, for net baryon density one has $[\cN]=3$ and hence $\Delta_{\cN}=3$. Since both temperature and baryon chemical potential have the mass dimension $[T]=[\mu]=1$, their conformal weight is $\Delta_T=\Delta_{\mu}=1$ .
Using analogous technique to that leading to Eq.~(\ref{eq:Urel}) the transformation rules needed to map the quantities expressed in de Sitter coordinates back to the polar Milne coordinates can be written as
\ba 
U_\mu(\tau,r) &=&\tau \frac{\partial \hat{x}^\nu}{\partial x^\mu}\,\hat{U}_\nu(\rho)\, , \nonumber \\
\cE(\tau,r) = \frac{\hat{\cE}(\rho)}{\tau^4} ,\,\,\,\, 
\cP(\tau,r) &=& \frac{\hat{\cP}(\rho)}{\tau^4} , \,\,\,\, 
\cN(\tau,r) = \frac{\hat{\cN}(\rho)}{\tau^3}\, , \nonumber \\ 
T(\tau,r) = \frac{\hat{T}(\rho)}{\tau},\,\,&&\,\,
\mu(\tau,r) = \frac{\hat{\mu}(\rho)}{\tau}.
\label{eq:trans}
\ea 

In \rf{eq:L}, $x^\b$ have mass dimension $[x^\b]=-1$ with one contravariant index, hence $\Delta_{x^\b}=0$, and using the information about the conformal weight of $\cE$ and $U^\a$ and \rf{eq:T}, one can find that the stress energy tensor should have the conformal weight $\Delta_{T^{\a\b}}=6$. As a result one observes that the conformal weight of the spin tensor is $\Delta_{S^{\a\b\g}}=6$, since each term in \rf{eq:L} should have the same conformal weight.
Hence the GLW spin tensor in \rf{eq:S} \cite{DeGroot:1980dk} as well as canonical spin tensor~\cite{Weinberg:1995mt}, and the Hilgevoord-Wouthuysen (HW) spin tensor~\cite{Speranza:2020ilk,Weickgenannt:2020aaf} should all have the same conformal weight. For instance, for a free Dirac field the canonical spin tensor $S_{\rm C}^{\a,\b\g}$, the GLW spin tensor $S_{\rm GLW}^{\a,\b\g}$ and the HW spin tensor $S_{\rm HW}^{\a,\b\g}$ are defined as~\cite{Speranza:2020ilk,Weickgenannt:2020aaf}
 \ba
 \label{eq:scan}
 S_{\rm C}^{\a,\b\g} &=& \frac{i}{8}\bar{\psi}\{\gamma^\a,[\gamma^\b,\gamma^\g]\}\psi\, ,\\
 \label{eq:GLW}
 S^{\alpha,\beta\gamma}_{\rm GLW}&=&\frac{i}{4m}\left(\bar{\psi}\sigma^{\beta\gamma}\olra{\partial}^\alpha\psi-\partial_\rho \epsilon^{\beta\gamma\alpha\rho}\bar{\psi} \gamma^5\psi \right),\\
 \label{eq:HW}
 S_{\rm HW}^{\a,\b\g} &=& S_{\rm C}^{\a,\b\g}\nonumber\\
 && -\frac{1}{4m}\left( \bar{\psi} \sigma^{\b\g} \sigma^{\a\rho}\partial_\rho \psi + \partial_\rho \bar{\psi} \sigma^{\a\rho} \sigma^{\b\g}\psi\right),
\ea
respectively, with
$\sigma^{\mu \nu}\!\!=\!\!\frac{i}{2}\left[\gamma^{\mu}, \gamma^{\nu}\right]
$. Since the conformal weight of both spinor $\psi$ and dual spinor $
\bar{\psi}\equiv\psi^{\dagger} \gamma_{0}
$ is $\Delta_\psi = \Delta_{\bar{\psi}} = \frac{3}{2}$ and conformal weight of Dirac gamma matrix is $\Delta_{\gamma^\mu}=1$~\cite{Kastrup:2008jn,Fabbri:2011ha}, therefore $\Delta_{S^{\a\b\g}}=\Delta_{S_{\rm C}^{\a\b\g}}=\Delta_{S_{\rm HW}^{\a\b\g}}=6$.

Similarly, using the information about the conformal weight of $\cN$ in \rf{eq:N} we find that the conformal weight of the net baryon number current is $\Delta_{N^\a}=4$.

The conformal weight of the spin polarization tensor $\omega^{\a\b}$ is easily found if one notices the fact that the first term in \rf{eq:S} should have the same conformal weight as the spin tensor in \rf{eq:L}. In this case for the rank-two dimensionless spin polarization tensor with two contravariant indices one can find the conformal weight to be $\Delta_{\omega^{\a\b}}\!=\!2$. Using this result in \rfm{eq:kappaomega} we see that $\Delta_{\kappa^{\a}}\!=\!1$ and $\Delta_{\omega^{\a}}\!=\!1$. Here we used the fact that $\epsilon^{\a\b\g\d}$ has zero mass dimension and four contravariant indices, giving $\Delta_{\epsilon^{\a\b\g\d}}\!=\!4$.
From the information of how $\kappa^{\a}$, $\omega^{\a}$ and basis vectors \refb{eq:pmbasis} transform under Weyl rescaling and that all the spin polarization components $a_i$ and $b_i$ are dimensionless scalars, we see that $\Delta_{a_i}\!=\!\Delta_{b_i}\!=\!0$. Hence, they are conformally invariant quantities.

Using \rf{eq:weyl-rescaling}  we summarize the transformation rules under Weyl rescaling (for four dimensional spacetime) as follows
\beq
\label{eq:ConfN}
N^\a &\rightarrow& \Omega^4 ~N^\a \, ,\\
\label{eq:ConfT}
T^{\a\b} &\rightarrow& \Omega^6 ~T^{\a\b} \, ,\\
\label{eq:ConfS}
S^{\a\b\g} &\rightarrow& \Omega^6 ~S^{\a\b\g}.
\eeq
Finally, we note that the similar results may be found for the quantities with covariant indices. In particular one may find that $\Delta_{T_{\a\b}}\!=\!2$. The latter result is in agreement with the fact that raising (lowering) the Lorentz index with the metric tensor changes the conformal measure by a factor of $2 \;(-2)$.
 
\section{Conformal invariance of conservation equations}
\label{sec:invariance}
%
Considering four-dimensional conformal fluid dynamics, we aim at finding the conformal transformation of the conservation equations for net baryon number, energy and linear momentum, and spin, which, in general coordinates, are expressed as
\ba
\label{eq:dN}
d_\a N^\a(x) &=& \partial_\a N^\a + \Gamma^\a_{\a\b} N^\b = 0\, , \\
\label{eq:dT}
d_\a T^{\a\b}(x) &=& \partial_\a T^{\a\b} + \Gamma^\a_{\a\lambda} T^{\lambda\b} + \Gamma^\b_{\a\lambda} T^{\a\lambda} = 0\, , \\
\label{eq:dS}
d_\a S^{\a\b\g}(x) &=& \partial_\a S^{\a\b\g}  \\ &&+ \nonumber\Gamma^\a_{\a\lambda} S^{\lambda\b\g} 
 + \Gamma^\b_{\a\lambda} S^{\a\lambda\g} + \Gamma^\g_{\a\lambda} S^{\a\b\lambda} = 0,
\ea
respectively, where $\Gamma^\b_{\a\lambda}$ are Christoffel symbols defined in \rf{eq:christoffel}.

To find the conformal transformations of the conservation laws \rfn{eq:dN}, \rfn{eq:dT}, and \rfn{eq:dS} we need to know the conformal transformation of the Christoffel symbols, which is given as~\cite{Wald:106274,Faraoni:1998qx,Loganayagam:2008is}
\beq
 \Gamma_{\lambda\a}^{\b}= \hat{\Gamma}_{\lambda\a}^{\b} + \delta^{\b}_{\lambda}\partial_{\a}\varphi+ \delta^{\b}_{\a}\partial_{\lambda}\varphi-
 \hat{g}_{\lambda\a}\hat{g}^{\b\sigma}\partial_{\sigma}\varphi \, ,
 \label{eq:dG}
\eeq
where $\delta^{\b}_{\lambda}$ is the Kronecker delta function and $\varphi$ is a function of space-time coordinates defined through \rf{eq:weyl-rescaling} (for the derivation of \rf{eq:dG} we refer the reader to App.~\ref{app:Confchristoffel}).

One can show, that for the four-dimensional spacetime the conservation law for net baryon number~(\ref{eq:dN}) is already conformal-frame independent~\cite{Baier:2007ix,Bhattacharyya:2007vs,Loganayagam:2008is,Gubser:2010ze,Gubser:2010ui}, i.e. net baryon number is conserved in both Minkowski and de Sitter space-times. In this case, since the conformal weight of net baryon number is $\Delta_{N^\a}=4$, one can write
\beq
d_\a N^{\a} =\Omega^4  \hat{d}_{\a}\hat{N}^{\a}.
\label{eq:ConsN}
\eeq
Using \rf{eq:ConfT} and \rf{eq:dG} in \rf{eq:dT}, we find that the conservation of energy and linear momentum transforms as~\cite{Wald:106274,Bhattacharyya:2007vs,Loganayagam:2008is}
\beq
d_\a T^{\a\b} =\Omega^6 \left[\hat{d}_{\a}\hat{T}^{\a\b}- \hat{T}^{\lambda}_{\HP\lambda}\hat{g}^{\b\delta}\partial_\delta\varphi\right]\,.
\label{eq:ConsT}
\eeq 
Hence, we observe that $\hat{T}^{\a\b}$ needs to be traceless in order to be conserved in de Sitter spacetime. Therefore, the breaking of conformal invariance is characterized only by the trace of the energy-momentum tensor~\footnote{Note that the latter holds regardless the symmetry of the energy-momentum tensor.}~\cite{Callan:1970ze,Wess:1971eb,DiFrancesco:1997nk,Forger:2003ut}.

Furthermore, using \rf{eq:ConfS} and \rf{eq:dG} in \rf{eq:dS} the conformal transformation of the conservation law for spin takes the form
\beq
d_\a S^{\a\b\g} &=& \Omega^6 \left[\hat{d}_{\a}\hat{S}^{\a\b\g}\right.\label{eq:ConsS}\\
&&-\left.(\hat{S}_{\lambda}^{\HP\lambda\g}\hat{g}^{\b\sigma} + \hat{S}^{\a\b}_{\HP\HP\a}\hat{g}^{\sigma\g})\partial_\sigma\varphi \right]. \nonumber
\eeq

From \rf{eq:ConsS} we find that the conformal invariance of the spin conservation law requires the spin tensor to satisfy the condition  $\hat{S}_{\a}^{\HP\a\b}=0$.
It is straightforward to show that the GLW (\ref{eq:S}, \ref{eq:GLW}) and HW \rfn{eq:HW} definitions do not satisfy this condition, and hence, explicitly break the conformal invariance of \rf{eq:ConsS}. The consequences of this issue will be discussed in the next section.
%
\section{Dynamics in de Sitter coordinates}
\label{sec:dynamics}
%
In this section we explore the dynamics of the spin polarization on top of the conformal Gubser-expanding hydrodynamic background in the de Sitter space-time. Our final results are obtained by transforming back to the Minkowski spacetime using \rfm{eq:trans}.
%
\subsection{Conformal symmetry breaking and the equation of state}
%
In Sec.~\ref{sec:invariance} we have shown that \rftwo{eq:Ncon}{eq:Tcon} are conformally invariant, i.e. they transform homogeneously under Weyl transformation provided the constitutive relations do so and the energy-momentum tensor is traceless. As the presence of finite masses breaks the tracelessness condition (in our case $\hat{T}^{\lambda}_{\HP\lambda} =3 \hat{\cP} - \hat{\cE}\approx {\cal O} (m^2)$), in order to respect Gubser symmetry and keep the four-velocity pattern invariant, the energy density \rfn{eq:calE}, pressure \rfn{eq:calP} and net baryon density \rfn{eq:calN} have to be treated in the massless limit, meaning
\beq
\hat{\cE}&=&24~k \ch{\alpha}~{\hat{T}}^4\,,
\label{eq:GE} \\
\hat{\cP}&=&8~k \ch{\alpha}~{\hat{T}}^4\,,
\label{eq:GP} \\
\hat{\cN}&=&8~k \sh{\alpha}~{\hat{T}}^3\,,
\label{eq:GN}
\eeq
respectively. Obviously, in this case one has $\hat{\cE}=3\hat{\cP}$.

On the other hand, for the conservation of spin to respect conformal invariance it is sufficient that the spin tensor in \rf{eq:ConsS} is totally antisymmetric~\footnote{Note that 
this is not the case for the GLW and HW definitions whose additional terms $\sim 1/m$ make them ill-defined in the massless limit as well as break the conformal invariance of the spin conservation.}. As noticed in  Sec.~\ref{sec:invariance} the GLW form \rfn{eq:S} of the spin tensor considered in this work does not satisfy this requirement. In general, if \rfm{eq:Ncon}, \rfn{eq:Tcon}, and \rfn{eq:Scon} were fully coupled this would lead to breaking of the Gubser symmetry and, in consequence, of the flow invariance. However, in the particular case studied herein, this is not the case since the spin dynamics given by \rf{eq:Scon} is treated only perturbatively~\cite{Florkowski:2019qdp}, and hence decouples from the background hydrodynamic fields. This allows for a separate solution of background equations of motion \rfn{eq:Ncon} and  \rfn{eq:Tcon} and subsequent solution of the spin part \rfn{eq:Scon}. As there is no back-reaction to the background from the evolution of spin dynamics (polarization tensor does not enter \rfm{eq:Ncon} and \rfn{eq:Tcon}), the conformal-breaking dynamics of spin polarization does not spoil the flow invariance (a similar issue was encountered also in other recent works, see for instance Ref.~\cite{Du:2020bxp}). At this point, it is important to mention, that, despite this issue, we are still allowed to investigate the spin dynamics in the de Sitter coordinates, which we will do for the sake of convenience. Using the same arguments as given above we will also keep finite masses in expressions for ${\cal A}$, ${\cal B}$, and ${\cal C}$ defining the GLW spin tensor in \rftwo{eq:S}{eq:SDelta}.
%
\subsection{Perfect-fluid background}
\label{ss:pfb}
%
Using \rf{eq:N} in \rf{eq:Ncon} the conservation law for charge in the de Sitter coordinates can be written as
\beq
\hat{U}^{\a}\p_{\a}\hat{\cN}+\hat{\cN}\p_{\a}\hat{U}^{\a} +\hat{\cN} ~\hat{U}^{\a} \frac{\p_\a \sqrt{-\hat{g}}}{\sqrt{-\hat{g}}}=0\, ,
\label{eq:chargedS}
\eeq
where $\hat{g}$ is the determinant of de Sitter metric~(\ref{eq:g}) and we employed definitions of covariant derivative from App.~\ref{app:covderiv}.
Identifying expansion scalar as $\hat{\theta}\equiv 2\tanh{\rho}$ one may rewrite \rf{eq:chargedS} in the following form
\beq
\p_\rho\hat{\cN}+\hat{\cN} \hat{\theta}=0\,.\lab{eq:charge}
\eeq
In the similar way, transforming \rf{eq:Tcon} to $dS_3{\otimes}{R}$, using \rf{eq:T}, and  contracting it with $\hat{U}_{\b}$, we get the conservation law for energy as follows
\beq
\p_\rho\hat{\cE}+(\hat{\cE}+\hat{\cP}) \hat{\theta}=0\, .\lab{eq:en}
\eeq
One may check that all other projections of \rf{eq:Tcon} are satisfied identically.

The solutions of \rf{eq:charge} and \rf{eq:en} can be found analytically giving~\cite{Gubser:2010ui,Gubser:2010ze},
\beq
\hat{\cE}&=&\hat{\cE}_0 \left(\frac{\cosh{\rho_0}}{\cosh{\rho}}\right)^{8/3},
\label{eq:GESol} \\
\hat{\cN}&=&\hat{\cN}_0 \left(\frac{\cosh{\rho_0}}{\cosh{\rho}}\right)^2,
\label{eq:GNSol}
\eeq
respectively, where $\hat{\cE}_0\equiv\hat{\cE}(\rho_0)$ and $\hat{\cN}_0\equiv\hat{\cN}(\rho_0)$ are integration constants and $\rho_0$ is the initial de Sitter time. Using \rfmtwo{eq:GE}{eq:GN} the corresponding solutions for  temperature and baryon chemical potential in de Sitter space are
\begin{figure}[t]
\begin{center}
\includegraphics[width=8.5cm]{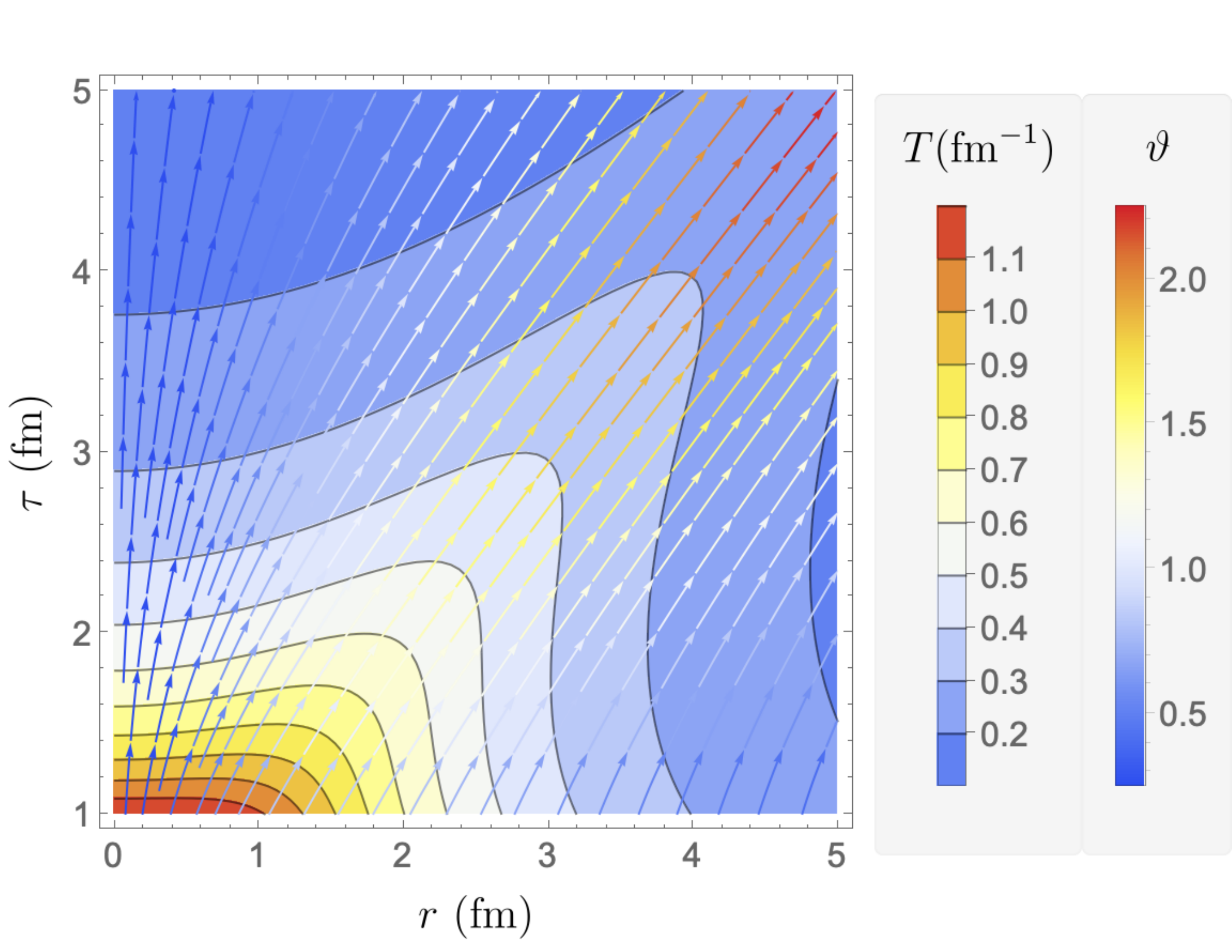}
\end{center}
\caption{\small The space-time dependence of temperature (contours) and flow-vector components $\left(U^{\tau}, U^{r}\right) / \sqrt{\left(U^{\tau}\right)^{2}+\left(U^{r}\right)^{2}}$ (stream lines -- the coloring of arrows is given by the rapidity $\vartheta$); see the bar legends for the scalings. }
\label{fig:TandU}
\end{figure}
%
\ba
\label{eq:GTSol}
\hat{T}&=&\hat{T}_0 \left(\frac{\cosh{\rho_0}}{\cosh{\rho}}\right)^{2/3}, \\
\label{eq:GMuSol}
\hat{\mu}&=&\hat{\mu}_0 \left(\frac{\cosh{\rho_0}}{\cosh{\rho}}\right)^{2/3}.
\ea
where, again, $\hat{T}_0\equiv\hat{T}(\rho_0)$ and $\hat{\mu}_0\equiv\hat{\mu}(\rho_0)$ are constants of integration. Based on above solutions we observe that $\hat{\alpha}$ is a $\rho$ independent quantity. 
Moreover, as shown in \cite{Gubser:2010ui,Gubser:2010ze} one can see that the Gubser symmetry restricts the dynamics of the system in such a way that it depends only on the de Sitter time $\rho$. 

The space-time evolution of temperature as given by~\rf{eq:GTSol} and flow-vector components $\left(U^{\tau}, U^{r}\right) / \sqrt{\left(U^{\tau}\right)^{2}+\left(U^{r}\right)^{2}}$ is shown in Fig.~\rfn{fig:TandU}. For setting initial temperature parameters we followed Ref.~\cite{Marrochio:2013wla} where it was assumed that $\hat{T}_0\equiv\hat{T}(\rho_0)=1.2$ at $\rho_0=0$ so that choosing $q\!=\!1 \,\rm{fm^{-1}}$ results in  \mbox{$T(\tau_0=1 \,{\rm fm}, r=0)=1.2 \,{\rm fm^{-1}}$}. One can observe that the temperature and flow profiles are strongly correlated.
%
\subsection{Spin evolution}
In this section we derive the spin evolution equations which we use to  study spin dynamics in the Gubser-flowing perfect-fluid background presented in the previous section. Herein, we assume that the spin polarization components, apart from $\rho$ dependence, also exhibit $\theta$ dependence. Hence, following the findings in Sec.~\ref{sec:invariance}, we assume that, in general, spin equations of motion break the conformal invariance. In such a  case $\rho$ and $\theta$ coordinates in the de Sitter space serve just as an alternative parameterization of the directions ($\tau,r$) in polar Milne coordinates.
By substituting \rf{eq:S} with \rf{eq:SDelta}, \rf{eq:k_decom} and \rf{eq:o_decom} in \rf{eq:Scon}, employing  \rfmtwo{eq:GTSol}{eq:GMuSol}, and subsequently projecting the resulting tensorial equation on $\hat{U}_\b \hat{R}_\g$, $\hat{U}_\b \hat{\Phi}_\g$, $\hat{U}_\b \hat{Z}_\g$,  $\hat{\Phi}_\b \hat{Z}_\g$, $\hat{R}_\b \hat{Z}_\g$ and $\hat{R}_\b \hat{\Phi}_\g$ we obtain the following equations of motion for the spin polarization components
\begin{widetext}
\ba
\hat{\cB}~\dot{\hat{a}}_R &=& -\hat{a}_R \left[\dot{\hat{\cB}} + \frac{5}{2} \hat{\cB} \tanh{\rho}\right] \, ,  \label{eq:ar}\\
\hat{\cB}~\dot{\hat{a}}_\Phi + \frac{\hat{\cB}}{2} \cosh{\rho} \sin{\theta} ~\mathring{\hat{b}}_Z &=& -\hat{a}_\Phi \left[\dot{\hat{\cB}} + \frac{5}{2} \hat{\cB} \tanh{\rho}\right]- \hat{b}_Z\frac{\hat{\cB}}{2}  \cosh{\rho} \cos{\theta} \, , \label{eq:apbz1}\\
\hat{\cB}~\dot{\hat{a}}_Z - \frac{\hat{\cB}}{2} \cosh{\rho} \sin{\theta} ~\mathring{\hat{b}}_\Phi &=& -\hat{a}_Z \left[\dot{\hat{\cB}} +\, 3 \,\hat{\cB} \tanh{\rho}\right]+ \hat{b}_\Phi\,\hat{\cB}\cosh{\rho}\cos{\theta}  \, , \label{eq:azbp1}\\
\left(\hat{\cB}-\hat{\cC}\right)~\dot{\hat{b}}_R &=& -\hat{b}_R \left[\left(\dot{\hat{\cB}}-\dot{\hat{\cC}}\right) + \left(\frac{9\hat{\cB}}{2}-4\hat{\cC}\right) \tanh{\rho}\right] \, , \label{eq:br}\\
\left(\hat{\cB}-\hat{\cC}\right)~\dot{\hat{b}}_\Phi -\frac{\hat{\cB}}{2\sin{\theta}}(\sech{\rho})^3 ~\mathring{\hat{a}}_Z &=& -\hat{b}_\Phi \left[\left(\dot{\hat{\cB}}-\dot{\hat{\cC}}\right) + \left(\frac{9\hat{\cB}}{2}-4\hat{\cC}\right) \tanh{\rho}\right] \, , \label{eq:azbp2}\\
\left(\hat{\cB}-\hat{\cC}\right)~\dot{\hat{b}}_Z +\frac{\hat{\cB}}{2\sin{\theta}}(\sech{\rho})^3 ~\mathring{\hat{a}}_\Phi &=& -\hat{b}_Z \left[\left(\dot{\hat{\cB}}-\dot{\hat{\cC}}\right) + \left(5\hat{\cB}-4\hat{\cC}\right) \tanh{\rho}\right] - \frac{\hat{\cB}}{2\sin{\theta}}\cot{\theta}(\sech{\rho})^3 ~\hat{a}_\Phi \, ,
\label{eq:apbz2}
 \ea
\end{widetext}
 respectively, where $\dot{(\HP)}$ $\equiv$ $\p_\rho$, $\mathring{(\HP)}$ $\equiv$ $\p_\theta$, and $\hat{\cB}$ and $\hat{\cC}$ are defined in \rf{eq:B} and \rf{eq:C}. In the latter expressions, unlike in the case of hydrodynamic background, we use full expressions for energy density \rfn{eq:calE} and pressure \rfn{eq:calP} expressed in terms of finite particle masses. 
%
\subsubsection{Massless limit}
\label{sss:massless}
From \rfmtwo{eq:ar}{eq:apbz2} we observe that, unlike in the case of Bjorken expansion~\cite{Florkowski:2019qdp}, only spin polarization components along $R^\mu$ evolve independently when expressed in de Sitter coordinates. On the other hand $\hat{a}_\Phi$ ($\hat{b}_\Phi$) and $\hat{b}_Z$ ($\hat{a}_Z$), respectively, are coupled to each other. The coupling between the components emerge due to the conformal symmetry breaking, and manifests itself through the $\theta$ dependence of the latter.

In the massless case the solutions to \rftwo{eq:ar}{eq:br} may be found analytically giving 
\ba
\label{eq:ar0}
\hat{a}_R&=&\hat{a}_R^0 \left(\frac{\cosh{\rho}}{\cosh{\rho_0}}\right)^{5/6}, \\
\label{eq:br0}
\hat{b}_R&=&\hat{b}_R^0 \left(\frac{\cosh{\rho_0}}{\cosh{\rho}}\right)^{7/6},
\ea
where, $\hat{a}_R^0\equiv\hat{a}_R(\rho_0)$ and $\hat{b}_R^0\equiv\hat{b}_R(\rho_0)$. The characteristic concave dependence of $\hat{b}_R$ on de Sitter time is qualitatively similar to that of temperature and baryon chemical potential; (in the case of $\hat{a}_R$ one deals with the convex function of $\rho$).

The dynamics of $\hat{b}_\Phi$ and $\hat{a}_Z$ components following from \rftwo{eq:azbp1}{eq:azbp2} is more complicated, however, it shows certain characteristic features. In particular, if $\hat{b}_\Phi$ is initially negligible, the  $\hat{a}_Z$ component is approximately $\theta$-independent, yielding
\ba
\label{eq:az0}
\hat{a}_Z&|_{\hat{b}_\Phi=0}=&\hat{a}_Z^0 \left(\frac{\cosh{\rho}}{\cosh{\rho_0}}\right)^{1/3},
\ea
with $\hat{a}_Z^0\equiv\hat{a}_Z(\rho_0)$, i.e. giving $\hat{a}_Z(\rho)\sim 1/\sqrt{\hat{T}(\rho)}$. 

In the general case where $\hat{b}_\Phi\neq 0$, one may use the fact that $\hat{a}_Z$ is a slowly varying function of $\theta$ and the second term on the left-hand side of \rf{eq:azbp2} may be neglected. In this case the solution to $\hat{b}_\Phi$ component has the approximate form
\ba
\label{eq:bP}
\hat{b}_\Phi&\approx&\hat{b}_\Phi^0 \left(\frac{\cosh{\rho_0}}{\cosh{\rho}}\right)^{7/6},
\ea
where $\hat{b}_\Phi^0\equiv\hat{b}_\Phi(\rho_0)$. Using exact numerical solutions one can show that, indeed, $\hat{b}_\Phi$ is a weakly-dependent function of $\theta$ and hence approximately proportional to $\hat{b}_R$.

In the case of the remaining $\hat{a}_\Phi$ and $\hat{b}_Z$ components simple solutions cannot be found and in general \rftwo{eq:apbz1}{eq:apbz2} have to be solved numerically. However, due to specific structure of \rftwo{eq:apbz1}{eq:apbz2} one may find some special solutions by requiring the $\theta$ terms to vanish, thus making the $\hat{a}_\Phi$ and $\hat{b}_Z$ independent of each other. The latter takes place for the following solutions
\ba
\label{eq:aP}
\hat{a}_\Phi&\approx& \hat{a}_\Phi^0  \left(\frac{\cosh{\rho}}{\cosh{\rho_0}}\right)^{5/6}\!\! \csc \theta,\\
\label{eq:bZ}
\hat{b}_Z&\approx&  \hat{b}_Z^0  \left(\frac{\cosh{\rho_0}}{\cosh{\rho}}\right)^{5/3}\!\! \csc \theta,
\ea
where, again,  $\hat{a}_\Phi^0\equiv\hat{a}_\Phi(\rho_0)$ and $\hat{b}_Z^0\equiv\hat{b}_Z(\rho_0)$, and hence $\hat{a}_\Phi\sim \hat{a}_R$. Finally, we observe that, in the cases discussed above all components exhibit the universal dependence of the type $(\cosh\rho)^{c}$, with $c$ being positive constant for $a_i$ and negative for $b_i$.
%
\subsubsection{Numerical results}
\label{sss:Results}

\begin{figure}[t]
\begin{center}
\includegraphics[width=8.5cm]{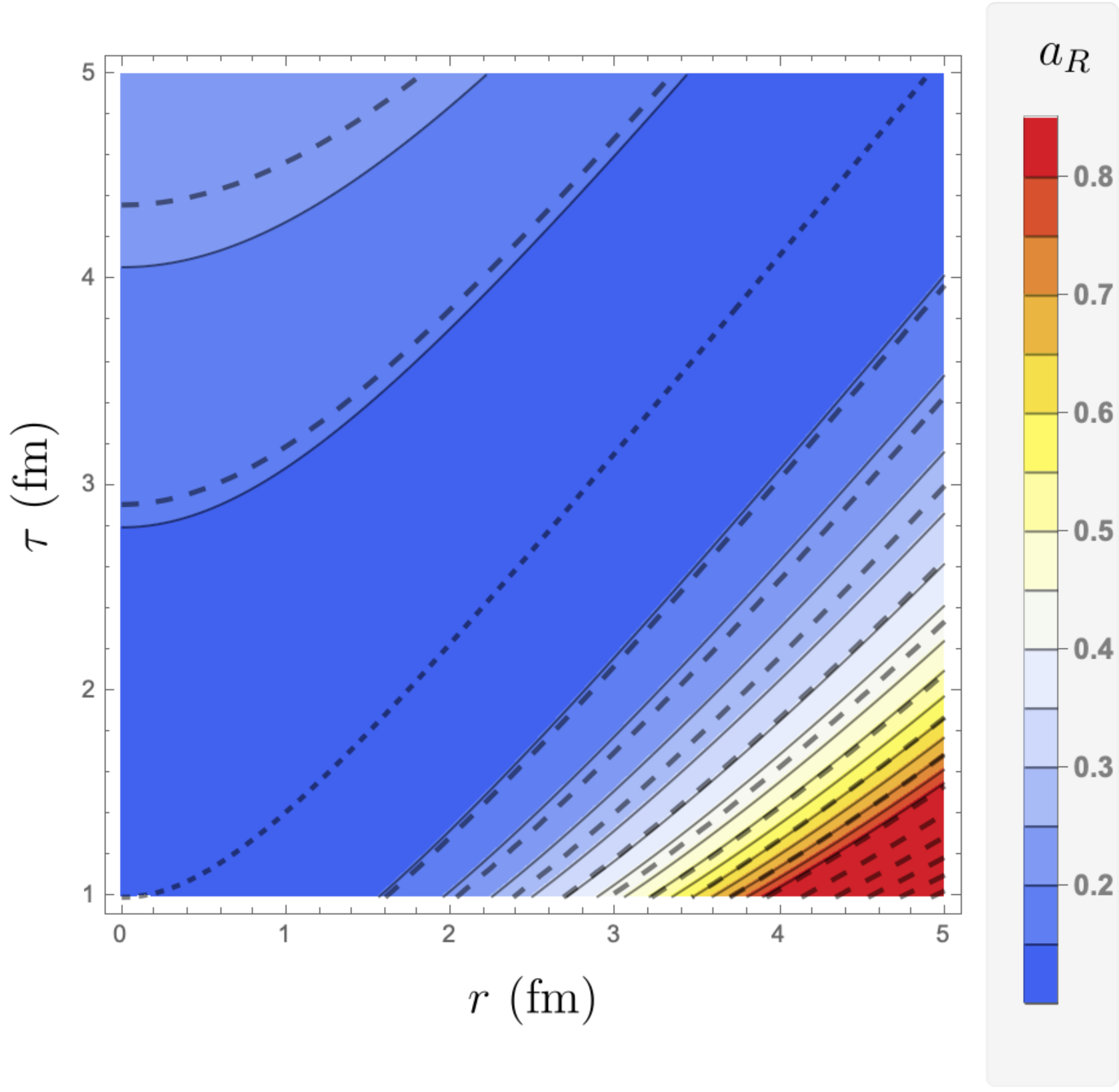}
\includegraphics[width=8.5cm]{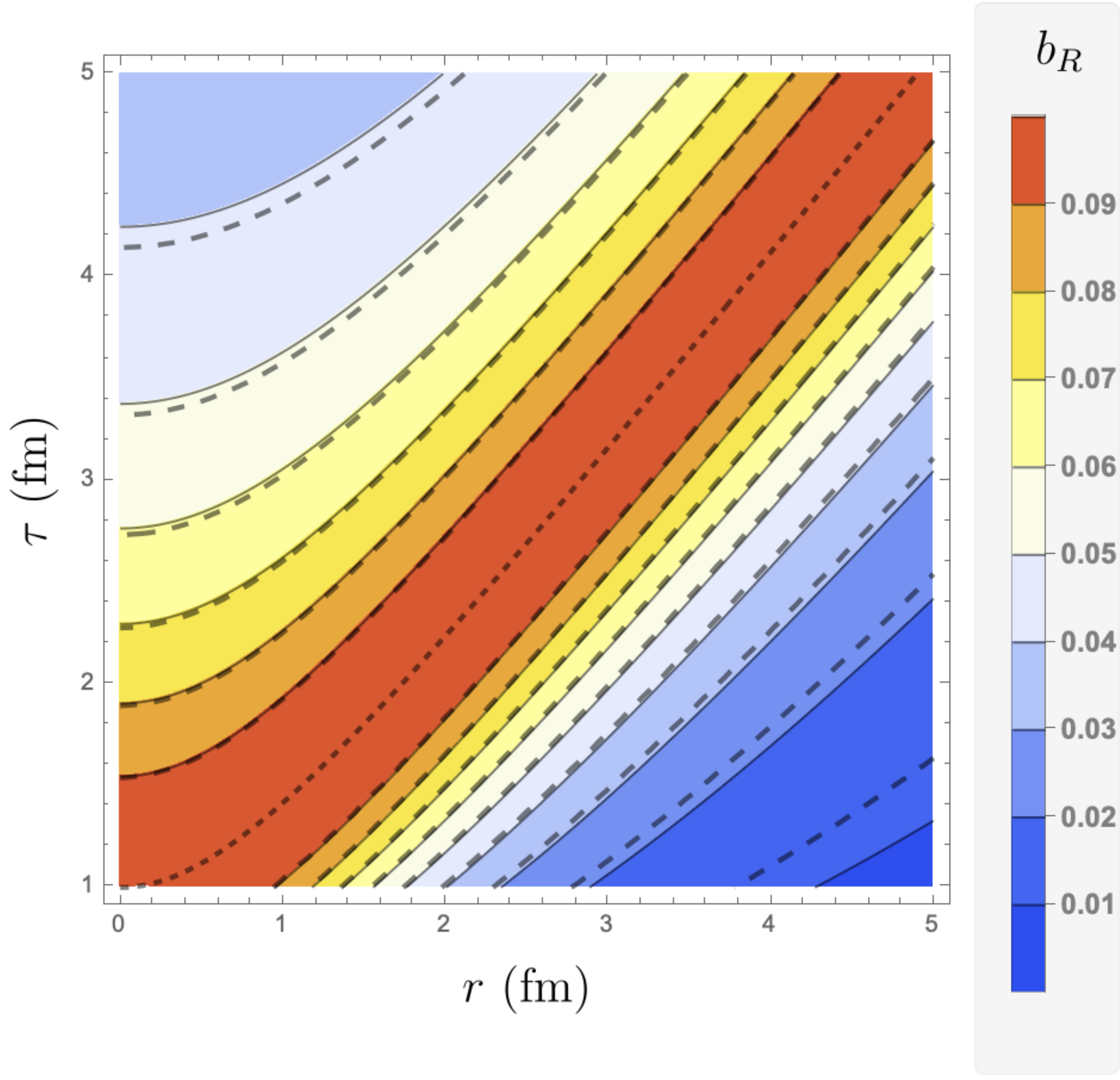}
\end{center}
\caption{\small Numerical solutions for $a_R$ and $b_R$ components of the spin polarization tensor as functions of proper time $\tau$ and radial distance $r$. }
\label{fig:arbr}
\end{figure}

\begin{figure}[t]
\begin{center}
\includegraphics[width=8.5cm]{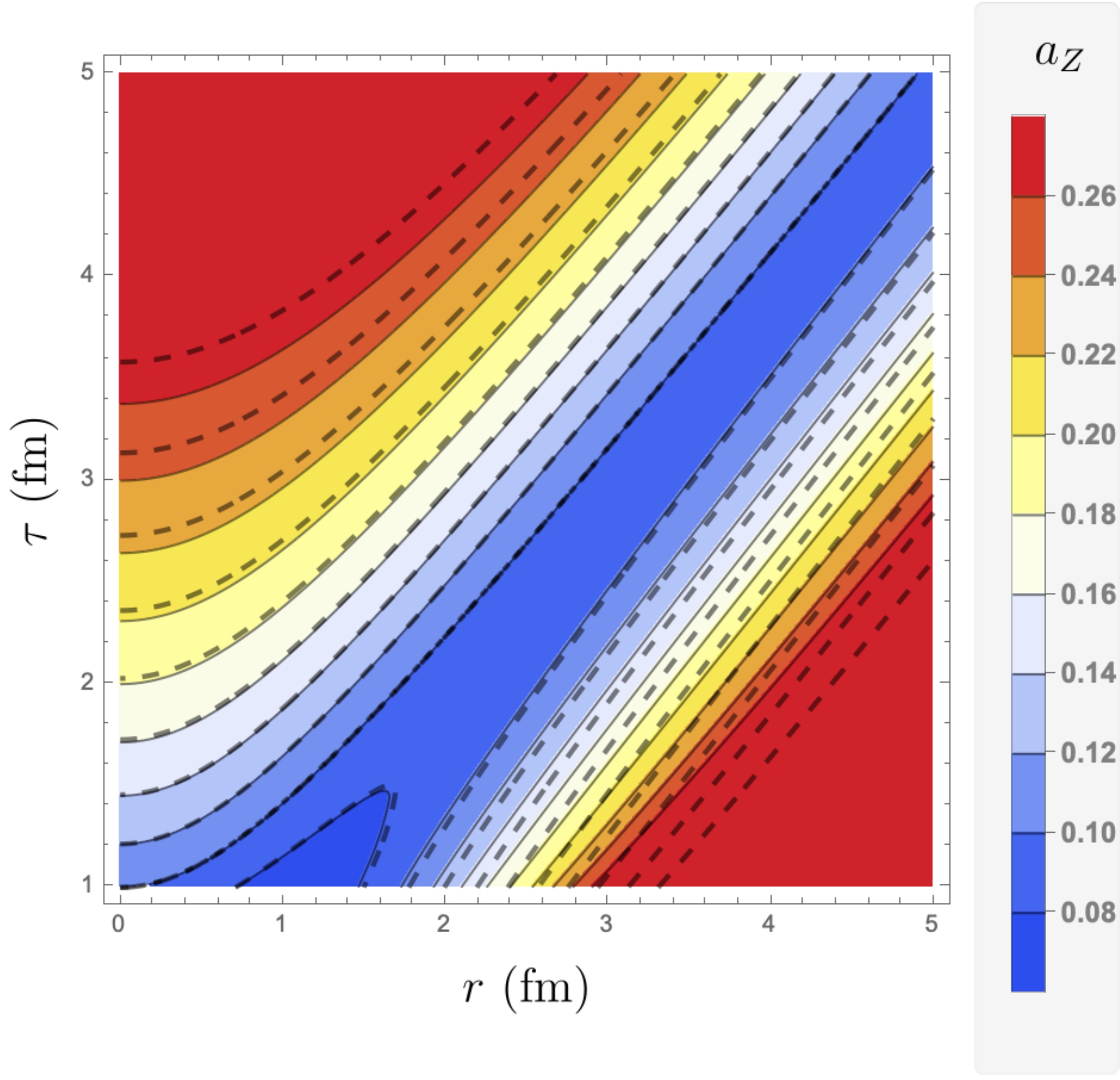}
\includegraphics[width=8.5cm]{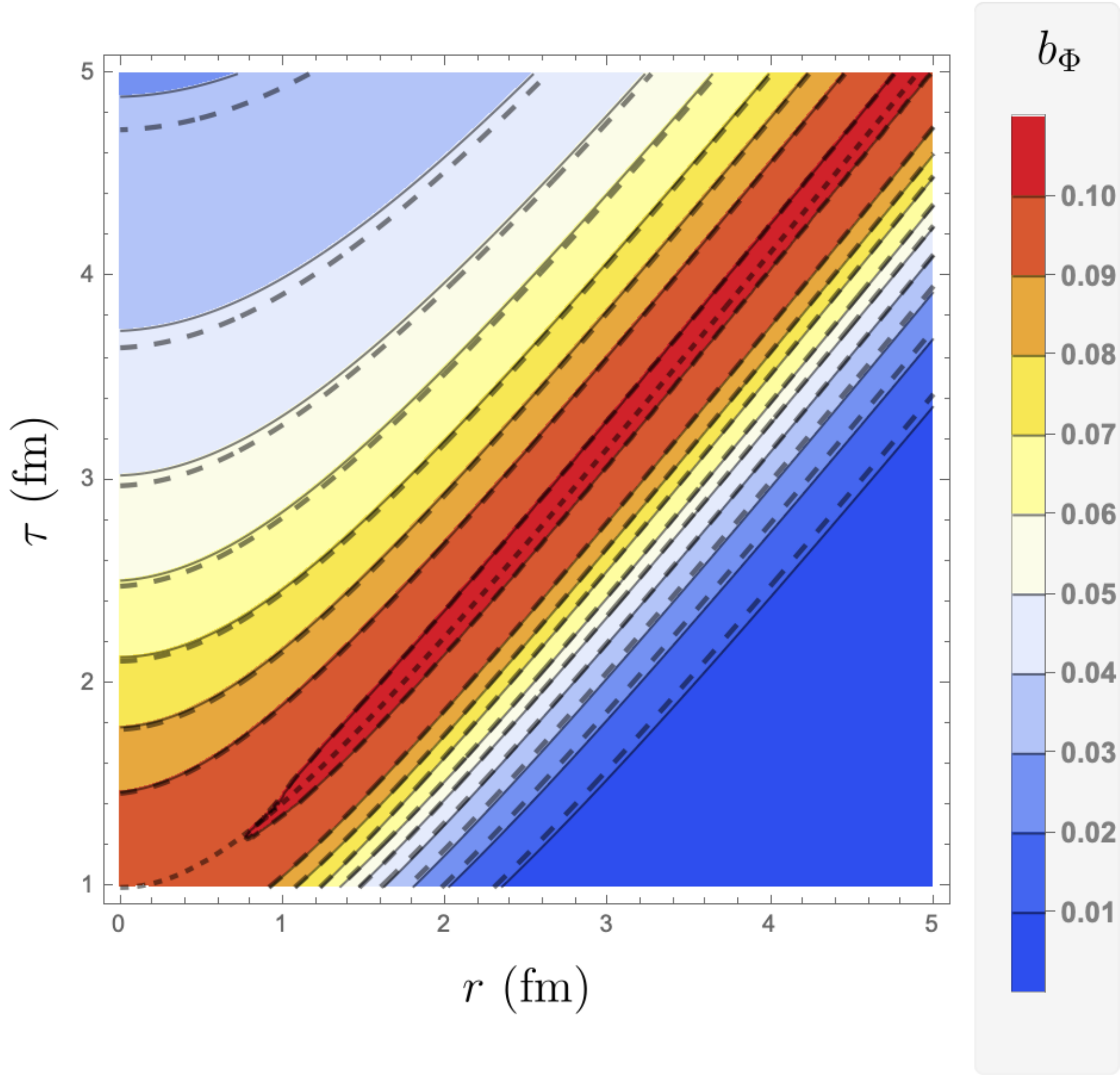}
\end{center}
\caption{\small Same as Fig.~\ref{fig:arbr} but for $a_Z$ and $b_\Phi$ components. }
\label{fig:azbp}
\end{figure}
 
Let us turn to some numerical solutions of the spin equations of motion. 
In Fig.~\rfn{fig:arbr} we present the exact numerical solutions of \rftwo{eq:ar}{eq:br} for $a_R$ and $b_R$ components as functions of proper time $\tau$ and radial distance $r$. The initial values of the $\hat{a}_R$ and $\hat{b}_R$ components are chosen in such a way so that $\hat{a}_R^0 =0.1$ and  $\hat{b}_R^0 =0.1$, which implies \mbox{$a_R(\tau_0=1 {\rm fm}, r=0)=b_R(\tau_0=1 {\rm fm}, r=0)=0.1$}. The mass in the calculations is set to  $m = 0.5 \hat{T}_0$. We find that the dynamics of $a_R$ and $b_R$ components is qualitatively different with $a_R$ having minimum and $b_R$ having a maximum at $\rho_0$ (the black dotted lines). Moreover, by comparing with the massless case we checked (see dashed lines) that the mass is weakly affecting the polarization dynamics as long as the mass is small, i.e. $z<1$).

In Fig.~\rfn{fig:azbp} we present analogous plots to Fig.~\rfn{fig:arbr} but for $a_Z$ and $b_\Phi$ components using the  numerical solutions of \rftwo{eq:azbp1}{eq:azbp2}. In this case we initialize the components in the similar way as previously assuming $\hat{a}_Z^0 =0.1$ and  $\hat{b}_\Phi^0 =0.1$. We observe a weak $\theta$ dependence of the solutions which is due to mixing of the two components. A comparison to massless case (dashed lines) confirms a small sensitivity of the solution to mass of the particles. Moreover we find that the increase of mass in general increases the values of the spin polarization components within the considered region. A similar conclusions hold for the remaining $a_\Phi$ and $b_Z$ components.
\section{Summary and conclusions}
\label{sec:summary}
In this paper we have used the formalism of hydrodynamics with spin~\cite{Florkowski:2018fap,Florkowski:2019qdp} to derive the equations of motion for spin polarization in the Gubser-expanding conformal hydrodynamic background~\cite{Gubser:2010ze,Gubser:2010ui}. Following previous works on this topic we have first solved the perfect-fluid hydrodynamical equations using the Gubser symmetry arguments in the de Sitter coordinates and obtained analytical solutions for hydrodynamic variables, see \rfmtwo{eq:GESol}{eq:GMuSol}. Subsequently, we have extended the analysis of the properties of the conservation laws with respect to the conformal  transformation  to the case of the angular momentum conservation. We have found that the latter is conformaly invariant provided the spin tensor is completely antisymmetric and the particles are massless.
As, in general, the currents in the GLW (Groot - van  Leeuwen - van Weert) framework studied here break these requirements we have solved the spin dynamics in the Gubser expanding background in the de Sitter coordinates (for the sake of convenience) allowing, however, for both, $\rho$ and $\theta$ dependence, of the spin polarization components. As there is no back reaction of the latter on the background we performed this analysis in the general massive case. The solutions obtained in this way are mapped back to the Milne coordinates, giving us the complete spatio-temporal evolution for the system with boost-invariance and cylindrical symmetry. In the massless limit we have found certain special solutions which, to a large extent, exhibit power-law type dependence on temperature. We have found that, unlike in the Bjorken case, only radial components behave independently. 

The framework used herein might be used, for instance, for the description of head-on collisions of initially polarized particles/ions at high energies~\cite{Fernow:1981fw,Milner:2013aua}, describing the mixing between polarization components along beam and in the azimuthal direction. Moreover, the formalism worked out here can be used for cross checking the numerical simulations within the full 3+1D geometry framework, which is now being developed.

\medskip
\section*{Acknowledgements}
We would like to thank Sayantani Bhattacharyya, Wojciech Florkowski, Masoud Shokri, David S\'en\'echal, and Yifan Wang for inspiring discussions and clarifications. R.S would like to acknowledge the kind hospitality during the school ``Frontiers in Nuclear and Hadronic Physics 2020'' at The Galileo Galilei Institute for Theoretical Physics in Florence, Italy where this work was initiated and also thank the Yukawa Institute for Theoretical Physics at Kyoto University where discussions
during the YITP workshop YITP-W-20-03 on ``Strings and Fields 2020'' were useful to complete this work. This research was supported in part by the Polish National Science Center Grants No. 2016/23/B/ST2/00717 and No. 2018/30/E/ST2/00432.
\appendix
\section{The covariant derivative}
\label{app:covderiv}
In a more general spacetime the partial derivative operator is not a good tensorial operator, therefore we would need the covariant derivative operator which is an extension of partial derivative to arbitrary manifolds, and which reduces to partial derivative operator in flat spacetime in Cartesian coordinates.
Its action on an arbitrary scalar $V$, rank-1 $V^\mu$ and rank-2 $V^{\mu\nu}$ tensors is
\ba
d_\mu V &=&\partial_\mu V\, , \\
\label{eq:covariant-derivative1} 
d_\mu V_\nu &=&\partial_\mu V_\nu-\Gamma^\sigma_{\mu\nu}V_\sigma\, , \\
\label{eq:covariant-derivative2}
d_\lambda V_{\mu\nu}&=&\partial_\lambda V_{\mu\nu}-\Gamma^\sigma_{\lambda\mu}V_{\sigma\nu}
-\Gamma^\sigma_{\lambda\nu}V_{\mu\sigma}\, ,\\
\label{eq:covariaCovariantnt-derivative3}
d_\mu V^\nu &=&\partial_\mu V^\nu+\Gamma^\nu_{\mu\sigma}V^\sigma\, , \\
\label{eq:covariant-derivative4}
d_\mu V^\mu &=&\partial_\mu V^\mu+\Gamma^\mu_{\mu\sigma}V^\sigma\, , \\
\label{eq:covariant-derivative5}
d_\mu V^{\mu\nu}&=&\partial_\mu V^{\mu\nu}+\Gamma^\mu_{\mu\sigma} V^{\sigma\nu}+\Gamma^\nu_{\mu\sigma} V^{\mu\sigma}\, ,\\
\label{eq:covariant-derivative6}
d_\lambda V^{\mu\nu}&=&\partial_\lambda V^{\mu\nu}+\Gamma^\mu_{\lambda\sigma} V^{\sigma\nu}+\Gamma^\nu_{\lambda\sigma} V^{\mu\sigma}\, ,
\label{eq:covariant-derivative7}
\ea
where $d$ denotes covariant derivative, $\partial$ is partial derivative, 
$g = \det{g_{\mu\nu}}$ and
$\Gamma^\nu_{\mu\lambda}$ are Christoffel symbols which are expressed as
\ba
\Gamma^\nu_{\mu\lambda}\equiv \Gamma^\nu_{\lambda\mu}=\frac{1}{2}g^{\nu\sigma}(\partial_\mu g_{\sigma\lambda}+\partial_\lambda g_{\sigma\mu}-\partial_\sigma g_{\mu\lambda})\, .
\label{eq:christoffel}
\ea
\section{Christoffel symbols in de Sitter coordinates}
\label{app:christoffel}
Starting from the definition of Christoffel symbol (\ref{eq:christoffel}) and using the de Sitter space-time metric (\ref{eq:deSitter-ds}), the following Christoffel symbols can be obtained which does not vanish
\ba
\Gamma^\rho_{\theta\theta}&=&\sinh\rho\cosh\rho\, , \\
\Gamma^\rho_{\phi\phi}&=&\sin^2\theta\sinh\rho\cosh\rho\, , \\
\Gamma^\theta_{\rho\theta}&=&\Gamma^\theta_{\theta\rho}=\tanh\rho\, , \\
\Gamma^\theta_{\phi\phi}&=&-\sin\theta\cos\theta\, , \\
\Gamma^\phi_{\rho\phi}&=&\Gamma^\phi_{\phi\rho}=\tanh\rho\, , \\
\Gamma^\phi_{\theta\phi}&=&\Gamma^\phi_{\phi\theta}=\cot\theta\, .
\ea
\section{Conformal transformation of Christoffel symbols}
\label{app:Confchristoffel}
In the de Sitter spacetime metric Christoffel symbol~\rf{eq:christoffel} is written as below where we use the transformation law of the metric tensor~\rf{eq:g-weyl} to obtain the conformal transformation of the Christoffel symbol
\ba
\hat{\Gamma}^\nu_{\mu\lambda}&=&\frac{1}{2}\hat{g}^{\nu\sigma}\left[\partial_\mu \hat{g}_{\sigma\lambda}+\partial_\lambda \hat{g}_{\sigma\mu}-\partial_\sigma \hat{g}_{\mu\lambda}\right]\, ,\nonumber\\
&& \text{Using Eq. (\ref{eq:g-weyl})},\nn\\
\hat{\Gamma}^\nu_{\mu\lambda}&=&\frac{1}{2}\Omega^{-2} g^{\nu\sigma}\left[\partial_\mu (\Omega^{2} g_{\sigma\lambda})+\partial_\lambda (\Omega^{2} g_{\sigma\mu})-\partial_\sigma (\Omega^{2} g_{\mu\lambda})\right]\, ,\nonumber\\
 &=&\frac{e^{2\varphi} g^{\nu\sigma}}{2} \left[\partial_\mu (e^{-2\varphi} g_{\sigma\lambda})+\partial_\lambda (e^{-2\varphi} g_{\sigma\mu})-\partial_\sigma (e^{-2\varphi} g_{\mu\lambda})\right]\, ,\nonumber\\
 &=&\frac{e^{2\varphi}g^{\nu\sigma}}{2}\left[ \vphantom{\int}  \right. 2 e^{-\varphi} \partial_\mu (e^{-\varphi}) g_{\sigma\lambda} + e^{-2\varphi} \partial_\mu (g_{\sigma\lambda})\nonumber\\
 && +2 e^{-\varphi} \partial_\lambda (e^{-\varphi}) g_{\sigma\mu} + e^{-2\varphi} \partial_\lambda (g_{\sigma\mu}) \nonumber\\
&&  -2 e^{-\varphi} \partial_\sigma (e^{-\varphi}) g_{\mu\lambda} - e^{-2\varphi} \partial_\sigma (g_{\mu\lambda})\left. \vphantom{\int}  \right]\, ,\nonumber\\
 &=& \frac{g^{\nu\sigma}}{2}\left[ \vphantom{int} \right. \partial_\mu (g_{\sigma\lambda}) +  \partial_\lambda (g_{\sigma\mu}) - \partial_\sigma (g_{\mu\lambda}) \nonumber\\
&& + e^{\varphi}\left( \partial_\mu (e^{-\varphi}) g_{\sigma\lambda}+ \partial_\lambda (e^{-\varphi}) g_{\sigma\mu}-\partial_\sigma (e^{-\varphi}) g_{\mu\lambda}\right)\left. \vphantom{int} \right]\, , \nonumber\\
 &=& \Gamma^\nu_{\mu\lambda} + e^{\varphi} \left[\delta^{\nu}_{\lambda} \partial_\mu (e^{-\varphi}) + \delta^{\nu}_{\mu} \partial_\lambda (e^{-\varphi})-\partial_\sigma (e^{-\varphi})g^{\nu\sigma} g_{\mu\lambda}\right]\, , \nonumber\\
\Gamma^\nu_{\mu\lambda}&=&\hat{\Gamma}^\nu_{\mu\lambda} + \delta^{\nu}_{\lambda} \partial_\mu \varphi + \delta^{\nu}_{\mu} \partial_\lambda \varphi-\hat{g}^{\nu\sigma} \hat{g}_{\mu\lambda} \partial_\sigma \varphi \, .\nn\\
\label{eq:conChristoffel}
\ea
Here we note  that for $\varphi(x)$ = \text{const.} one has $\hat{\Gamma}^\nu_{\mu\lambda}$ = $\Gamma^\nu_{\mu\lambda}$.

\bibliography{pv_ref}{}
\bibliographystyle{utphys}
\end{document}